\documentclass[12pt]{utarticle}   
\usepackage{latexsym,amsmath,amssymb,amscd,cite}
\usepackage[mathscr]{eucal}
\usepackage[dvips]{color}
\usepackage{rotating}
\DeclareFontFamily{U}{rsf}{}
\DeclareFontShape{U}{rsf}{m}{n}{
  <5> <6> rsfs5 <7> <8> <9> rsfs7 <10-> rsfs10}{}
\DeclareMathAlphabet\Scr{U}{rsf}{m}{n}

\newcommand{\al}{\alpha}

\newcommand{\de}{\delta}
\newcommand{\e}{\epsilon}

\newcommand{\la}{\lambda}

\newcommand{\s}{\sigma}

\newcommand{\om}{\omega}

\newcommand{\Ga}{\Gamma}

\newcommand{\Om}{\Omega}
\newcommand{\Si}{\Sigma}

\newcommand {\cA} {{\cal A}}

\newcommand {\cE} {{\cal E}}
\newcommand {\cF} {{\cal F}}

\newcommand {\cH} {{\cal H}}

\newcommand {\cL} {{\cal L}}

\newcommand {\cO} {{\cal O}}

\newcommand {\cQ} {{\cal Q}}

\newcommand {\cV} {{\cal V}}

\newcommand {\bbC} {\mathbb{C}}

\newcommand {\bbR} {\mathbb{R}}

\newcommand {\bbZ} {\mathbb{Z}}

\newcommand{\ba}{{\bar a}}
\newcommand{\bb}{{\bar b}}
\newcommand{\bi}{{\bar \imath}}
\newcommand{\bj}{{\bar \jmath}}

\newcommand{\bz}{{\bar z}}

\newcommand{\id}{{\rm id}}

\newcommand{\im}{{\rm im}}

\newcommand{\dl} {\partial}

\newcommand{\rarrow}{\rightarrow}

\newcommand{\ie}{i.e.~}

\newcommand{\End}{{\rm End}}

\long\def\beq #1 \eeq{\begin{equation}#1\end{equation}}
\long\def\bea #1 \eea{\begin{eqnarray}#1\end{eqnarray}}
\long\def\beann #1 \eeann{\begin{eqnarray*}#1\end{eqnarray*}}

\newcommand{\bit}{\begin{itemize}}
\newcommand{\eit}{\end{itemize}}

\newcommand{\ben}{\begin{enumerate}}
\newcommand{\een}{\end{enumerate}}

\long\def\bec #1 \eec{\begin{center}#1\end{center}}
\newtheorem{theorem}{Theorem}

\long\def\bth #1 \eth{\begin{theorem}#1~\end{theorem}}
\long\def\bpr #1 \epr{~\\ \emph{Proof:}~ #1~$\Box$}

\long\def\del#1\enddel{}
\long\def\new#1\endnew{{\bf #1}}

\def\nn{\nonumber{}}

\newcommand{\tfor}{{\qquad \mathrm{for} \quad}}

\newcommand{\tand}{{\quad \mathrm{and} \qquad}}

\newcommand{\twith}{{\qquad \mathrm{with} \qquad}}

\newcommand{\mapshort}[1]{\begin{picture}(30,20)(0,20)
    \put(8,30){{\small $#1$}}
    \put(0,25){\vector(1,0){30}}    
  \end{picture}
}

\newcommand{\tpsi}{\widetilde{\psi}}
\newcommand{\poisson}{\pi}%{\widetilde{\om}}
\newcommand{\fol}{\mathscr{F}}
\newcommand{\coiso}{C}

\newcommand{\cstr}{{\hat J}}

\newcommand{\tom}{\hat\om}

\begin{document}        
%\preprint{
%  CERN--PH--TH/2009-XXX\\}
  
\title{On higher rank coisotropic A-branes}

\author{
  Manfred Herbst
}
   \oneaddress{
     University~of~Augsburg,\\
     Institute~for~Mathematics,\\
     Universit\"atsstrasse~14,\\
     86159~Augsburg,~Germany\\
}

\Abstract{
  This article is devoted to a world sheet analysis of A-type D-branes 
  in $N=(2,2)$ supersymmetric non-linear sigma models. In addition to the
  familiar Lagrangian submanifolds with flat connection we reproduce the rank one
  A-branes of Kapustin and Orlov, which are supported on coisotropic submanifolds.
  The main focus is however on gauge fields of higher rank and on tachyon
  profiles on brane-antibrane pairs. This will lead to the notion of a complex of
  coisotropic A-branes. A particular role is
  played by the noncommutative geometry on the brane world volume. 
  It ensures that brane-antibrane pairs 
  localize again on coisotropic submanifolds.
}

\maketitle
\thispagestyle{empty}

\

---------

{\scriptsize 2010 PACS classification: 11.25.-w Strings and branes}

\newpage

{\small\tableofcontents}
%\newpage

\section{Introduction and summary}

Almost a decade ago Kapustin and Orlov realized in \cite{KO2001} that Lagrangian submanifolds with flat connection do not always exhaust all objects in the category of A-branes 
on a Calabi--Yau manifold $M$. Their argument in favour of new objects involved mirror symmetry together with K-theoretic considerations, and once convinced that the new objects are there, they employed string theoretic techniques to study their geometry in the context of $N=(2,2)$ non-linear sigma models. Their important insight was that an A-brane can be supported on a coisotropic submanifold $C$ of $M$ provided that there exits a line bundle over $C$ whose connection induces a transverse complex structure, a notion that will be defined in due time.

Ever since the trigger was set in \cite{KO2001} numerous developments sharpened the understanding of coisotropic A-branes. The new class of objects found applications in many interesting fields of mathematical physics, for instance in the geometric Langlands program \cite{KW2006}, in a new approach to quantization \cite{GW2008}, or in four-dimensional gauge theories \cite{NW2010}. Many of their properties will be reviewed as we go on. However, one that will not be elaborated on in this work, although important, is the extension of the notion of a stable A-brane from Lagrangian to coisotropic submanifolds \cite{KL2003,Li2004}.
 
An important question that led to some speculations, see for instance the concluding discussions in \cite{KO2003}, is the question of how to describe coisotropic A-branes with higher rank vector bundles. The methods that were applied in \cite{KO2001} and in the subsequent literature did not allow to go beyond rank one gauge groups, essentially because the D-brane geometry was extracted from the 
\emph{boundary conditions on the $N=(2,2)$ supercurrents}.
This work takes a different route and applies the methods that were successful in classifying B-type boundary conditions in Landau--Ginzburg models \cite{KL2002,BHLS2003,Lazaroiu2003} and in non-linear sigma models \cite{Hori2000,HHP2008}: The A-type supersymmetry will be checked directly on an $N=1$ supersymmetric action for the sigma model including boundary terms that encode the D-brane data. This approach will improve our understanding of the category of A-branes but, as we will see at the end, will also leave many questions unanswered and raise many new ones.

Section~\ref{GeometryofAbranes} starts with the analysis of rank one A-branes in $N=(2,2)$ non-linear sigma models on a (real) $2n$-dimensional Calabi--Yau manifold $M$.
This will reproduces the following two by now classical results from \cite{KO2001}.
\begin{enumerate}
  \item[I.] A-branes are supported on coisotropic submanifolds $C$ in $M$, \ie
        the Poisson bivector $\pi = \omega^{-1}$, associated with the K\"ahler form
        $\omega$, vanishes on the conormal bundle $N^*C$. 
\end{enumerate}
Standard symplectic geometry says that every coisotropic submanifold 
carries a \emph{characteristic foliation} ${\fol}$ 
induced from its conormal bundle via the Poisson bivector. More precisely,
vectors in $\pi^\#(N^*C) \subset TC$ can be integrated to the leaves 
of the foliation. Bearing this structure in mind, the second condition on the
geometry of A-branes is as follows.
\begin{enumerate}
  \item[II.] An A-brane carries a line bundle whose connection, together with
        the B-field, gives rise to a \emph{transverse complex structure} 
        $\cstr = \pi \cF$, that is a complex structure
        on the transverse bundle $N\fol$ of the characteristic foliation. Here, 
        $\cF = B+F$ is a combination of the B-field and the curvature of the gauge
        connection. $\cstr$ turns out to restrict the dimension of $C$ to be $n + 2k$ for 
        $k$ a non-negative integer.
\end{enumerate}

Before proceeding with the outline, let us pause for a brief discussion on the existence of a transverse complex structure $\cstr$. As noted in \cite{KO2003} the condition $\cstr^2 = -\id$ can be reformulated in a more intuitive way as
$$
  (\omega - i\cF)^{\wedge k}|_{N\fol} \neq 0, \tand
  (\omega - i\cF)^{\wedge (k\!+\!1)}|_{N\fol} = 0.
$$
Notice that for A-branes on Lagrangian submanifolds the condition is void. From the effective field theory point of view it can be interpreted as F-term condition \cite{FIM2006}, which obstructs deformations of the K\"ahler parameters in the presence of a coisotropic A-brane with given curvature $F$. Thus, already at the classical level coisotropic A-branes can only exist at holomorphic submanifolds of the complexified K\"ahler moduli space, an effect that is unfamiliar from Lagrangian A-branes. 
Note however that such situations are familiar from B-branes in the mirror dual picture. There, it may happen that a B-brane obstructs deformations of the complex structure of $M$. Explicit examples were constructed using matrix factorization techniques in \cite{HW2004}.

In Section~\ref{subsec:topologicaltwist} we proceed by topologically twisting the theory in order to get control over two important effects on coisotropic A-branes.
\begin{enumerate}
  \item[III.] The first is \emph{perturbative} in nature and concerns the
        \emph{noncommutative product} on $C$. It is non-trivial only 
        transverse to the characteristic
        foliation, and it is actually transversely holomorphic. More precisely, 
        the holomorphic Poisson bivector that controls the product is the
        $(\widehat{2,0})$-part of $\pi$ with respect to $\cstr$.
  \item[IV.] The second is a \emph{non-perturbative} effect from
        instantons, \ie from holomorphic maps of the world sheet with boundary into $M$, 
        where each boundary component is mapped to a single leaf of the 
        characteristic foliation. 
\end{enumerate}
In \cite{Kapustin2003,Kapustin2005} generalized complex geometry already indicated that coisotropic A-branes should carry a holomorphic noncommutative geometry.
It was made manifest in the seminal work on the geometric Langlands program \cite{KW2006}, where on specific hyper-K\"ahler manifolds a canonical coisotropic A-brane gave rise to a description of A-branes in terms of D-modules (that is modules over the sheaf of differential operators).

With the preparations of Section~\ref{GeometryofAbranes} at hand we can attack the actual objective of this work in Section~\ref{AandT}, that is to find a more general form of A-branes, including stacks of coisotropic A-branes and tachyon profiles between brane--antibrane pairs. To this end we split off a trace-part from the gauge field that, together with the B-field, determines the boundary conditions of the non-linear sigma model, thus leading to the structures I -- IV. The remaining non-Abelian part of the gauge field is put into a path-ordered exponential of an $N=1$ superconnection, which can be further extended to include a tachyon profile for brane--antibrane pairs \cite{Hori2000,KL2000,TTU2000}.
$N=2_A$ supersymmetry then requires that
a stack of A-branes on a coisotropic submanifold $C$ with fixed transverse complex structure $\cstr$ is given by a complex of vector bundles with the following properties:
\begin{enumerate}
  \item[V.] The bundles in the complex are 
        \emph{noncommutative transversely holomorphic vector 
        bundles}. More precisely, given the characterisitc foliation $\fol$ on $C$ 
        the connections of the vector bundles have, firstly, a vanishing curvature
        along the leaves of $\fol$ and, secondly, in transverse direction a curvature 
        of type $(\widehat{1,1})$ with respect to $\cstr$.
  \item[VI.] The differential $Q$ in the complex is 
        \emph{noncommutative transversely holomorphic}. In
        particular, it is constant along the leaves, and the condition to be a
        differential involves the noncommutative product, that is $Q*Q=0$.
\end{enumerate}
As we will see in Section~\ref{subsec:roleNCG}, the noncommutative geometry induced by the $(\widehat{2,0})$-part of $\pi$ plays an important role for the consistency of the tachyon condensation process. It ensures that the complex localizes again on coisotropic submanifolds. In particular, the minimal dimension accessible by tachyon
condensation is that of a Lagrangian submanifold.

The link of higher-dimensional coisotropic A-branes to Lagrangian ones via tachyon condensation suggests a relation between holomorphic noncommutative geometry and disc instanton corrections. The idea that such a link should exist goes back to \cite{BS2002}, however, the role of coisotropic A-branes seems to be new and will be discussed in Section~\ref{sec:lessons}, where we also draw some conclusions and close with a discussion on missing steps towards a proper categorical description of A-branes.

\subsubsection*{Acknowledgements}
I thank Anton Kapustin, Nikita Nekrasov and Edward Witten for valuable discussions and comments. This research was supported by the ERC Starting Independent Researcher Grant StG No. 204757-TQFT.

\section{The geometry of A-branes}
\label{GeometryofAbranes}

Let us consider an $N=(2,2)$ supersymmetric non-linear sigma model, mapping a two-dimensional world sheet $\Si$ into a $2n$-dimensional K\"ahler manifold $M$ with a complex structure $J$, a hermitian metric $g$ and a K\"ahler form $\om$. The latter is complexified by the B-field $B$. We assume $M$ to be a Calabi--Yau manifold in order to ensure an unbroken axial R-symmetry, which will be important in Section~\ref{AandT}.

We consider the theory on a strip with coordinates $(\s^0,\s^1) \in \bbR \times [0,\pi]$ and flat metric with signature $(-1,1)$.
The field content of the non-linear sigma model is given by chiral (and anti-chiral) multiplets, whose component fields are the holomorphic coordinate fields $x^i$ (and $\bar x^\bi$) on $M$ and their superpartners $\psi^i_\pm$ (and $\bar\psi^\bi_\pm$). 
For the subsequent discussion of A-type boundary conditions, it will be convenient to work in real coordinates, $x^I = (x^i,x^\bi)$ and 
$\psi^I_\pm = (\psi^i_\pm,\bar\psi^\bi_\pm)$. Let us introduce 
\beq
  \nonumber %\label{N1coordinates}
  \psi^I := \psi^I_+ + \psi^I_- , \tand \quad
  \tpsi^I := \psi^I_+ - \psi^I_- .
\eeq

\subsection{Boundary conditions with $N=1$ world sheet supersymmetry}

Our strategy to find general A-type boundary conditions is to start with an $N=1$ supersymmetric boundary action and impose $N=2_A$ invariance. 
A manifestly $N=2$ supersymmetric description of (rank one) D-branes in general $N=(2,2)$ theories including besides chiral superfields also twisted chiral and semi-chiral superfields can be found in \cite{SSW2009}.
 
In this subsection we introduce the $N=1$ quantities that are relevant for the subsequent discussions.
  
The unbroken $N=1$ subalgebra of $N=(2,2)$ is the one lying in both, the supersymmetry preserved by A-branes and the one preserved by B-branes. It acts on the component fields as
\bea
  % \label{N1algebra}
  \nn \de_1 x^I &=& i \e_1 \psi^I, \\
  \nn \de_1 \psi^I &=& -2 \e_1 \dl_0 x^I, \\
  \nn \de_1 \tpsi_I &=& - 2 \e_1 g_{IJ} \dl_1 x^J,
\eea
where $\tpsi_I = g_{IJ} \tpsi^J$. 

The following action consists of the one for the non-linear sigma model (cf. for instance \cite{MirrorBook}), which is $N=(2,2)$ supersymmetric up to boundary terms, and a boundary action that ensures total $N=1$ invariance,
$$
  S = S_{nlsm} + \int \!d\s^0~ \frac i4 \psi^I (g_{IJ}\tpsi^J - B_{IJ}\psi^J)  \Big|_0^\pi\ .
$$
We may introduce a gauge field $A$, which we assume to be rank one for the moment, so that the field strength is $F = dA$. The corresponding $N=1$ boundary action is \cite{HIV2000}
$$
  S_A = \int d\s^0 \left( A_I \dl_0 x^I - \frac i4 F_{IJ}\psi^I  \psi^J \right) \Big|_0^\pi .
$$

The general variation of this action leads to $N=1$ supersymmetric boundary conditions that restrict the string end points to a submanifold $\coiso$ in $M$, \ie for all $t \in T\coiso$ and $n \in N^*\coiso$ we require
\beq
  \label{N1conditions}
    n_I \de x^I = n_I \dl_0 x^I = n_I \psi^I = 0, \tand
    t^I N^b_I = t^I N^f_I = 0 ,
\eeq
where $N^b_I := g_{IJ} \dl_1 x^J - \cF_{IJ} \dl_0 x^J$ and 
$N^f_I := g_{IJ} \tpsi^J - \cF_{IJ} \psi^J$. Having said this we can properly interpret $A$ as the connection of a line bundle over $C$. $\cF=B+F$ is the gauge invariant combination of the B-field and the curvature $F$.

\subsection{A-branes as coisotropic submanifolds}

Taking the $N=1$ action as starting point we now determine the consequences of $N=2_A$ supersymmetry on the geometry of the submanifold $C$ and on the gauge field $A$. 

The variations of the component fields of the chiral multiplets are
\bea
  \label{NAalgebra}
  \nn \de_A x^I &=& i \e_A J^I{}_J \tpsi^J, \\
      \de_A \psi_I &=& -2 \e_A \om_{IJ} \dl_1 x^J, \\
  \nn \de_A \tpsi^I &=& 2 \e_A J^I{}_J \dl_0 x^J.
\eea
Applying them to the above action gives
\bea
  \label{bdryAvar}
  \nn \de_A (S+S_A) = \int d\s^0 ~i \e_A ~
  \bigg(\hspace*{-20pt}&& ~\poisson^{IJ} N^b_I N^f_J + \\
      &&+~ \frac i2 \psi^I \dl_I 
         \left(N^f_K ~\poisson^{KL}\cF_{LJ} \right) \psi^J
         - \frac i2 \psi^I N_K^f \dl_I \poisson^{KL} \cF_{LJ}~ \psi^J \\
      \nn 
      &&+~ (\om_{IJ} + \cF_{IK}\poisson^{KL}\cF_{LJ})~ \dl_0 x^I \psi^J
  \bigg)\bigg|_0^\pi \ ,
\eea
which needs to vanish for A-type boundary conditions. 
Here the Poisson structure $\pi$ is the inverse of the symplectic structure on $M$. 

\subsubsection*{Coisotropic submanifolds}

\newcommand{\codim}{\mathrm{codim}\ }

The first line in (\ref{bdryAvar}) tells us that the Poisson structure $\poisson$ must vanish on the conormal bundle $N^*\coiso$, \ie for any pair of covectors $u$ and $v$ in $N_x^*\coiso$ at a point $x \in \coiso$ it requires
\beq
  \label{Defcoiso}
  \poisson^{IJ} u_I v_J = 0 \ .
\eeq
This is the defining property of a coisotropic submanifold. Its codimension $m$ can take values in $\{0,\ldots,n\}$. If $\coiso$ is half-dimensional, $m=n$, it becomes a Lagrangian submanifold.

Let us review the geometry of coisotropic submanifold in more detail. 
From the defining condition (\ref{Defcoiso}) we find that the Poisson structure along $\coiso$ induces a map $\poisson^\#: v_I \mapsto \poisson^{IJ}v_J$ from the conormal bundle $N^*\coiso$ to the tangent bundle $T\coiso$. The image of this map is the bundle of $\om$-orthogonal vectors to $T\coiso$,
$$
  T\coiso^\om = \{v \in T M|_\coiso~|~ \om_{IJ} v^I w^J=0 ~~
  \mathrm{for~all}~~ w \in T\coiso \}.
$$
In fact, the property  $T\coiso^\om \subset T\coiso$ serves as an alternative definition of a coisotropic submanifold. In view of the isomorphism
$$
  \poisson^\#: N^*\coiso \rightarrow T\coiso^\om,
$$
there is a relation of dimensions, $\dim T\coiso^\om = \codim C = m$.

\subsubsection*{The characteristic foliation}

The subbundle $T\coiso^\om \subset TC$ is \emph{integrable}, \ie the Lie bracket of any two vectors in $T\coiso^\om$ is again a vector in $T\coiso^\om$. This follows from $d\om=0$. Indeed, take $z\in T\coiso$ and $u,v \in T\coiso^\om$, then
$$
  0 = u^I v^J z^K \dl_{[I} \om_{JK]} = (u^I\dl_I v^J - v^I\dl_I u^J)z^K \om_{JK}.
$$
The Frobenius theorem then tells us that we can locally choose coordinates $y^1,\ldots,y^m$, so that $\{\dl/\dl y^1,\ldots, \dl/\dl y^m\}$ serves as a basis for $T\coiso^\om$, and that we can integrate along these vectors to obtain the \emph{characteristic foliation} $\fol$ of the coisotropic submanifold $\coiso$. Its leaves, $L$, are $m$-dimensional and have the tangent bundle $TL = T\coiso^\om|_L$.
The bundle $T\coiso^\om$ is therefore the tangent bundle to the foliation $\fol$, and we denote it henceforth by $T\fol$, a standard notation in foliation theory. 
Note that by definition the symplectic structure degenerates along the leaves of the characteristic foliation. 
For a Lagrangian submanifold $\coiso$ we have $T\fol = T\coiso$, and the foliation consists of only one leaf $L=\coiso$.

The transverse bundle to the foliation is defined as a quotient bundle,
$$
  N\fol := T\coiso/T\fol.
$$ 
However, the metric on the K\"ahler manifolds $M$ can be used to pick a canonical representative for the quotient, $N\fol \cong (T\fol)^\perp$. 
Note that the restriction of $\omega$ to $C$ is non-degenerate transverse to the foliation, that is on $N\fol$.

The complex structure $J$ induces an isomorphism 
$J : T\fol \rightarrow N\coiso$, and its restriction to the transverse bundle, 
$J : N\fol \rightarrow N\fol$ induces the decomposition 
$N\fol \otimes \bbC = N^{(1,0)}\fol \oplus N^{(0,1)}\fol$.
In fact, the coisotropic submanifold inherits a transverse K\"ahler structure $(\coiso,g,J,\om)$ from the ambient space \cite{Oh2003}, \ie $\coiso$ carries all the properties of a K\"ahler manifold but only on the transverse bundle $N\fol$.

\subsubsection*{The transverse complex structure}

Let us proceed with the second line of (\ref{bdryAvar}). In view of the isomorphism $\poisson^\#:N^*\coiso\rightarrow T\fol$, it requires
$t^I \cF_{IJ} = 0$ for any vector $t \in T_x\fol$ at a point $x\in \coiso$,
\ie $\cF$ is transversely polarised,%
\footnote{
The second line of (\ref{bdryAvar}) is actually weaker. The condition of $\cF$ being transversely polarised also follows from the $N=2_A$ supersymmetry of the boundary conditions (\ref{N1conditions}).
} 
$$
  \cF \in \Ga(\coiso,\wedge^2 N\fol)\ .
$$
On compact manifolds we have $B \in H^2(M,\bbR/\bbZ)$ and $F \in H^2(\coiso,\bbZ)$ and therefore we can 
set $B|_{T\fol} = 0$ and $F|_{T\fol}= 0$ independently.
For a Lagrangian submanifold this results in $B = F =0$ on $\coiso$ \cite{HIV2000}.

For the interpretation of the last line in (\ref{bdryAvar}) we follow reference \cite{KO2001} and define
$$
  \cstr^I{}_J := \poisson^{IK} \cF_{KJ} : N\fol \longrightarrow N\fol.
$$
$\cstr$ is non-trivial only in the transverse direction of the foliation, since both, $\om$ and $\cF$, are transverse on $C$. 
The last line in (\ref{bdryAvar}) then tells us that $\cstr$ is an almost complex structure on the transverse bundle $N\fol$, 
$$
  \cstr^2 = - \id .
$$
In fact, Kapustin and Orlov have shown in \cite{KO2001} that $\cstr$ is integrable. It therefore defines a \emph{transverse complex structure} on the foliation $\fol$ and induces a decomposition
$$
  N\fol \otimes \bbC = N^{(\widehat{1,0})}\fol \oplus N^{(\widehat{0,1})}\fol .
$$

It is easy to check that the two-forms $\cF$ and $\om$ are of type 
$(\widehat{2,0}) \oplus (\widehat{0,2})$ with respect to this decomposition. Antisymmetry and non-degeneracy of $\cF$ and $\om$ on $N\fol$ then require that the dimension of $N^{(\widehat{1,0})}\fol$ must be even, so that $2(n-m) = 4k$ for some integer $k$. This restricts the possible dimensions of the coisotropic submanifold,
$$
  \dim C = 2n-m = n + 2k \tfor k = 0,1,\ldots,\left[\frac n2 \right].
$$
The brackets denote taking the integer part.

The condition $\cstr^2 = -\id$ can be rewritten as a simple condition on the complexified K\"ahler class. To see this let us consider the operators $\hat P_\pm = 1/2 (\id \mp i \cstr)$ on $N\fol$. $\hat P_\pm$ are projection operators (to $N^{(\widehat{1,0})}\fol$ resp. $N^{(\widehat{0,1})}\fol$) if and only if $\cstr^2 = -\id$. 
Introducing the closed two-form
$$
  2\om \hat P_+ = \om - i \cF, %= (\om-iB) - iF ,
$$
we find that $\cstr$ is an almost complex structure if and only if the matrix $(\om-i\cF)$ has rank $2k$, or equivalently \cite{KO2003}
$$
  (\om - i\cF)^{\wedge (k+1)}|_{N\fol} = 0 , \quad \tand 
  (\om - i\cF)^{\wedge k}|_{N\fol} \neq 0.
$$
In the context of type II string compactifications this condition can be interpreted as F-term condition for the low-energy effective action \cite{FIM2006}. It is void for Lagrangian submanifolds, but gives an obstruction to deforming complexified K\"ahler moduli in the presence of higher dimensional coisotropic A-branes.

The property that $\om$ and $\cF$ are of type $(\widehat{2,0}) \oplus (\widehat{0,2})$, together with their relation to the projectors $\hat P_\pm$, tells us that $\om-i\cF$ (resp. $\om+i\cF$) is of type $(\widehat{2,0})$ (resp. $(\widehat{0,2})$).
In local $\cstr$-complex coordinates in transverse direction, say $z^a$ for $a = 0,1,\ldots, 2k$, we have
\beq
  \label{HatDecomposition}
  \begin{array}{rcl}
    \om_{ab}-i\cF_{ab} &=& 2 \om_{ab} = -2 i \cF_{ab},\\
    \om_{\ba\bb}+i\cF_{\ba\bb} &=& 2 \om_{\ba\bb} = 2 i \cF_{\ba\bb}.
  \end{array}
\eeq

\subsubsection*{A different real basis for the fermions}

Let us analyze the boundary conditions for coisotropic A-branes in another real description of the $N=(2,2)$ non-linear sigma model, one that fits better to the $N=2_A$ subalgebra. In fact, the linear combinations of fermions,
$\psi^I = \psi^I_+ + \psi^I_-$ and $\tpsi^I = \psi^I_+ - \psi^I_-$, were well-adapted to the two-step analysis above, where we started with the $N=1$ subalgebra. The drawback is however their non-homogeneous transformation under the axial $U(1)$ R-symmetry. To see this recall that the fermions carry axial R-charges according to the following table:
$$
  \begin{array}{|c||c|c|c|c|}
    \hline
        & \psi^i_+ & \psi^i_- & \psi^\bi_+ & \psi^\bi_- \\[2pt]
    \hline
    R_A &   +1    &   -1     &   -1       &   +1\\
    \hline
  \end{array}
$$
The fermions $\psi_+^I$ and $\psi_-^I$ are not Eigenvectors of the R-symmetry action, which suggests to use a different combination of fermions, namely
\beq
  \label{N2Acoordinates}
  \begin{array}{rcl}
    \chi^I &:=& (\psi^i_-,\bar\psi^\bi_+) \twith R_A(\chi^I) = -1 ,\\[2pt]
    \rho^I &:=& (\psi^i_+,\bar\psi^\bi_-) \twith R_A(\rho^I) = +1 .
  \end{array}
\eeq
These are related to the $N=1$ fermions by $\psi^I = \chi^I + \rho^I$ and 
$\tpsi^I = i J^I_{~K} (\chi^K - \rho^K)$. 

Let us consider the boundary conditions for a coisotropic A-brane in terms of $\chi^I$ and $\rho^I$. Along the coisotropic submanifold $C$ we had 
$N_I^f|_{N\fol} = (g_{IJ}\tpsi^J -\cF_{IJ}\psi^J)|_{N\fol}=0$, which becomes
$$
  (\om_{IJ}+i\cF_{IJ})\rho^J\big|_{N\fol} = (\om_{IJ}-i\cF_{IJ})\chi^J\big|_{N\fol} .
$$
Notice the appearance of the antiholomorphic and holomorphic projectors on the left- and right-hand side, respectively. As the two sides transform differently under R-symmetry rotations, they must vanish independently, that is
\beq
  \label{N2Afermcondition}
  (\hat P_+ \chi)^I\big|_{N\fol} = 0\ , \tand
  (\hat P_- \rho)^I \big|_{N\fol} = 0\ .
\eeq
We find that transverse to the foliation the fermions, $\rho$ and $\chi$, take values in $N^{(\widehat{1,0})}\fol$ resp. $N^{(\widehat{0,1})}\fol$.
The boundary conditions (\ref{N1conditions}) along $NC$ and $T\fol$ both imply
\beq
  \label{normalfermcondition}
  \chi^I\big|_{NC} 
  = \rho^I\big|_{NC}
  = 0\ .
\eeq

Condition (\ref{N2Afermcondition}) actually requires that the $(2,0)\oplus (0,2)$ part of $F$ with respect to the complex structure $J$ from $M$ must not vanish on a coisotropic A-brane $C$. To see this, notice that boundary conditions on fermions
relate left-movers, $\psi_+$, to right-movers, $\psi_-$. Since both $\om$ and $B$ have type $(1,1)$, it follows immediately from (\ref{N2Afermcondition}) that the $(2,0)\oplus (0,2)$ part of $F$ must be non-vanishing. On the other hand, the $(1,1)$ part of $F$ may vanish or cancel the $B$-field.  
We will say more about the latter situation in the next subsection.

Let us summarize what we found so far. 

\

\parbox{15cm}{
\emph{A rank one A-brane on a $2n$-dimensional K\"ahler manifold $(M,g,J,\om)$ with closed $B$-field of type $(1,1)$ is a coisotropic submanifold $C$ of dimension $n+2k$ together with a line bundle with connection $A$, whose curvature, $F=dA$, gives rise to a complex structure 
$\cstr = \poisson (B+F)$ on the transverse bundle of the 
characteristic foliation $\fol$.\\ 
The K\"ahler structure on $M$ induces a transverse K\"ahler structure on $C$. The $(2,0)\oplus(0,2)$ part of $F$ with respect to the complex structure $J$ on $M$ must be non-vanishing.}
}

\

\noindent After twisting to the topological A-model, one may generalize from K\"ahler to arbitrary symplectic manifolds $M$. Then the summary remains true except for the last two statements.

\subsection{Transverse K\"ahler and hyper-K\"ahler}
\label{subsec:hyperKahler}

Let us find a hermitian metric for the transverse complex structure $\cstr$. From the context of noncommutative geometry on the world volume of a D-brane \cite{SW1999} it is known that open strings do not couple to the metric $g$ on $M$, but rather to the boundary metric
$$
  G_{IJ} = g_{IJ} - \cF_{IK} g^{KL} \cF_{LJ}.
$$
On a coisotropic A-brane this metric reduces to $g$ along the leaves of the foliation, whereas it differs from $g$ in the transverse direction.
It is actually straight forward to see that
$$
  G_{KL} \cstr^K{}_I \cstr^L{}_J = G_{IJ} \quad\mathrm{on}\quad N\fol.
$$
Having identified $G$ as the hermitian metric we define its transverse K\"ahler form
$$
  \tom_{IJ} := -G_{IK} \cstr^K_{~J} = 
  \cF_{IK} J^K_{~J} + \cF_{KJ} J^K_{~I}\quad\mathrm{on}\quad N\fol.
$$
Notice that it is of type $(2,0)\oplus (0,2)$ with respect to $J$. In fact, we have 
$\tom_{ij} = 2i \cF_{ij}$ and $\tom_{\bi\bj} = -2i\cF_{\bi\bj}$.

It is natural to ask when the metric $G$ is transverse K\"ahler, that is when $d\hat\om=0$. To answer this question let us work in the complex coordinates of $M$. Since $\cF$ is closed we find 
$(d\tom)_{[\bi i j]} = \dl_\bi \tom_{jk} = 
  2i \dl_\bi \cF_{jk} = 
  2i \dl_{[k}\cF_{j]\bi}$. 
and hence we have the following:

\

\parbox{15cm}{
\emph{$(C,G,\cstr,\tom)$ is transverse K\"ahler if and only if the $(2,0)$ part of $\cF$ is $\bar\dl$-closed, or equivalently if the $(1,1)$-part of $\cF$ is $\dl$-closed,}
\beq
  \label{Kahler}
  d \tom = 0 
  \quad\Leftrightarrow\quad \bar\dl \cF_{2,0} = 0
  \quad\Leftrightarrow\quad \dl \cF_{1,1} = 0 .
\eeq
}

\

Thus, assuming (\ref{Kahler}) leads to two transverse K\"ahler structures on the coisotropic A-brane, the first being induced from the embedding space $(M,g,J,\om)$, and the second, $(C,G,\cstr,\tom)$, being induced by the gauge field on $C$. They come however with two different metrics. 

An obvious question is to ask under which circumstances the two metrics agree (up to a multiplicative constant) and thus define a transverse hyper-K\"ahler structure. 
As we found earlier the boundary conditions on the fermions require the $(2,0)$ part of $\cF$ to be non-vanishing. We might try however to set the $(1,1)$ part to zero, that is to require $\cF_{IK}J^K_{~J} = \cF_{KJ}J^K_{~I}$. In that case the condition to have an almost complex structure, $\cstr^2 = -\id$, is equivalent to
\beq
  \label{hyperKahler}
  G_{IJ} = 2g_{IJ} = - 2\cF_{IK} g^{KL} \cF_{LJ} \quad\mathrm{on}~~N\fol.
\eeq
We conclude that 

\

\parbox{15cm}{
\emph{$(C,g,J,\om,\cstr,1/2\,\tom)$ is transverse hyper-K\"ahler if and only if $\cF_{(1,1)}=0$.} 
}

\

\noindent The third transverse complex structure is 
$K = J\cstr = -\cstr J$ and its K\"ahler form is $\cF$.

\subsection{Topological twisting}
\label{subsec:topologicaltwist}

The real formulation (\ref{N2Acoordinates}) is especially well-adapted for twisting the $N=2_A$ invariant theory to the topological A-model on $M$. After Wick rotation, $\s^2=i\s^0$ and $z=\s^2-i\s^1$, and twisting, the fermions take values in $\chi^I \in x^*T(M)$ as well as
$\rho^i \in x^*T^{(1,0)}(M)\otimes K$ and 
$\rho^\bi \in x^*T^{(0,1)}(M)\otimes \bar K$, where $K$ is the canonical bundle on the world sheet. The BRST operator $Q$ acts as follows on the fields,
\beq
  \label{topalgebra}
  \begin{array}{cclcccl}
    Q x^I &=& \chi^I, \\
    Q \chi^I &=& 0\ ,  \\
    Q \rho^I &=& - (1 + i J)^I_{~K}\dl_z x^K -
    (1 - i J)^I_{~K}\bar\dl_\bz x^K  - F^I .\\
  \end{array}
\eeq
In the zero mode sector it is represented by the de-Rham differential $d$ on forms 
$\al = \al_{I_1\ldots I_p} dx^{I_1} \ldots dx^{I_p}\in \Om^p(M)$, \ie for an operator $O_\al = \al_{I_1\ldots I_p} \chi^{I_1} \ldots \chi^{I_p}$ we have
$Q O_\al ~=~ O_{d\al}$.
 
The topological observables $O_\al$ are in one-to-one correspondence with elements in the de-Rham cohomology $H^*(M)$, where the form-degree is the axial R-charge (or ghost number) of the corresponing operator.

\subsubsection*{Topological observables on coisotropic A-branes}

To see the action of the BRST operator $Q$ on the world sheet boundary recall the boundary conditions for the fermions, $\chi^a=0$ on $N\fol$ and $\chi^I=0$ on $NC$. Comparing these with the topological algebra (\ref{topalgebra}), the BRST operator is found to act trivially along $N^{(\widehat{1,0})}$, so that on $C$ it is represented 
by \cite{KL2005}
$$
  d_C := d_\parallel + \bar\dl_\perp = 
           dy^c \frac{\dl}{\dl y^c} + d\bz^\ba\frac{\dl}{\dl \bz^\ba},
$$
where $d_\parallel$ is the de-Rham differential along the leaves of the foliation $\fol$, and $\bar\dl_\perp$ is the Dolbeault differential transverse to the foliation.
Note that the latter is defined with respect to the transverse complex structure $\cstr$.

In order to describe boundary observables on a single coisotropic A-brane in the topological A-model we define 
$$
  \hat\Om(C) := T^*C^\bbC / (N^{(\widehat{1,0})}\fol)^* 
  \cong T^*\fol^\bbC \oplus (N^{(\widehat{0,1})}\fol)^*,
$$
and $\hat\Om^p(C) := \wedge^p \hat\Om(C)$. The 
topological observables on a rank one coisotropic A-brane are then in one-to-one correspondence with elements in
$$
  \hat H^p(C) := 
    \frac{\ker(d_C:\hat\Om^p    (C) \rarrow \hat\Om^{p+1}(C))}
         {\im (d_C:\hat\Om^{p-1}(C) \rarrow \hat\Om^p    (C))}
         \ .
$$

\subsubsection*{Deformations of coisotropic A-branes}

Elements in $\hat H^1(C)$ are infinitesimal deformations of the A-brane geometry. 
In fact, an infinitesimal deformations along a section $n \in NC$ leaves $C$ coisotropic if  $\cL_n \om= d \iota_n \om = 0$. To link these deformations to elements in $\hat H^1(C)$ recall that the symplectic form induces an isomorphism 
$NC \rarrow T^*\fol$, $n \mapsto \iota_n \om$, and so we have
$
  \cL_n \om = d_\parallel \iota_n \om = d_C \iota_n \om \in \wedge^2 T^*\fol 
$.
Therefore, $C$ stays coisotropic if and only if $d_C \iota_n \om = 0$. If we mod out by Hamilonian deformations, $\iota_n\om = d_\parallel f$, we find that the cohomology class $\iota_n \om \in \hat H^1(\fol,\bbR)$ (which is polarized along the leaves of $\fol$) defines a deformation of the coisotropic submanifold. The deformations $a \in \hat H^1(\fol,\bbR/\bbZ)$ of the gauge field along the foliation complexify $\iota_n \om$. 

For another example of infinitesimal deformations take a one-form $\hat a \in (N^{(\widehat{0,1})}\fol)^*$. It is a topological observable if it defines a $\bar\dl_\perp$ cohomology element. This corresponds to an infinitesimal 
deformation of the transverse complex structure (or to a deformation of a transversely holomorphic vector bundle, a notion that we will introduce in Section~\ref{holomorphicgaugefield}).

This discussion might be compared with \cite{KM2006}, where infinitesimal deformations of D-branes were studied in the context of generalized complex geometries.
In the mathematical literature, the deformation theory of coisotropic submanifolds was worked out by Oh and Park in \cite{OP2003}. On the other hand, the deformation theory of transversely holomorphic foliations was studied quite some time ago by Duchamp and Kalka \cite{DK1979} as well as G\'omez--Mont \cite{Gomez1980}. On A-branes both deformation theories need to be combined. However, as we will see later on, the noncommutative geometry on the coisotropic A-brane needs to be taken into account as well.

Finally, topological observables in $\hat H^0(C)$ are functions on $C$ with $d_C f = 0$, \ie $f$ has to be constant along the leaves and $\cstr$-holomorphic transverse to the foliation. 
Deformations by such transversely holomorphic functions, $f(z^a)$, have the interpretation of turning on a tachyonic field on $C$. This will be the main theme in Section~\ref{tachyon}.

\subsection{World sheet instantons for coisotropic A-branes}

World sheet instanton corrections play a major role in the topological A-model. The instanton equation,
$$
  (\id - i J)^I_{~K}\bar\dl_\bz x^K = 0,
$$
requires a holomorphic embedding of, say a disk, $D$ into $M$. The boundary of the disk, $\dl D$, has to be mapped into $C$ with some restrictions. To see them (cf. \cite{KW2006}) let us compare the Wick rotated version of the boundary conditions (\ref{N1conditions}) with the instanton equation. Let $\dl_1$ and $\dl_2$ be normal resp. tangent to the disk. In transverse direction to the foliation we have  
\beq
  \label{instantonbdry}
  0 = g_{IJ} \dl_1 x^J - i\cF_{IJ} \dl_2 x^J =
  (\om - i\cF)_{IJ} \dl_2 x^J ~~ \mathrm{on}~~ N\fol\ ,
\eeq
which implies that a tangent vector to $x(\dl D)$ cannot lie in $N\fol$.
On the other hand, the boundary condition along the leaves of the foliation is trivially satisfied by the instanton equation,
$$
  g_{IJ} \dl_1 x^J|_{T\fol} = \om_{IJ} \dl_2 x^J|_{T\fol} = 0.
$$
We have found the following:

\

\parbox{15cm}{
\emph{The boundary of a holomorphic disk instanton must be mapped into a single leaf $L$ of the characteristic foliation of $C$, $x:(D,\dl D) \rarrow (M,L)$. For space-filling coisotropic A-branes there are no instanton corrections, because the leaves are points.}
}

\

\

The world sheet action in the topologically twisted theory can be recast as
\beq
  \label{topolaction}
  S = \int_D x^*(\om - i B) - i \int_{\dl D} x^* A + \{Q,V\} .
\eeq
The topological terms depend on the complexified K\"ahler form and on the gauge field on $\coiso$. A deformation of the coisotropic submanifold by a normal vector $n \in NC$ can be included via
$\int_{\dl D} n^I g_{IJ} \dl_1 x^J$ plus terms to ensure BRST invariance.
On a world sheet instanton the action then becomes
$$
  S_{inst} = \int_ D x^*(\om - i B) + \int_{\dl D} x^* (\iota_n \om -i A).
$$
Since the instanton is sensitive only to the gauge field in leaf direction, $A_\parallel$, the instanton corrections in the path integrale are weighted by the holonomy of $A_\parallel+i\iota_n\om$ around the boundary circle $\partial D$.

\subsection{Holomorphic noncommutative geometry in the topological sector}

\newcommand{\PB}{{{}_{PB}}}

The ordering of observables on the world sheet boundary together with the B-field and the gauge field gives rise to a noncommutative deformation of the D-brane geometry \cite{Schomerus1999,SW1999}. In some situations this deformation does not play an essential role. For instance, for B-branes it does not affect the topological sector, because the latter is governed by holomorphic quantities whereas $\cF$ and the associated Poisson structure $\theta$ both are of type $(1,1)$. Also, for A-branes on Lagrangian submanifolds there is no effect, because $\om = \cF = 0$ thereon. Coisotropic A-branes are different, for them noncommutative geometry plays an essential role in the topologically twisted theory.
 
The bivector that gives rise to a noncommutative product was found in \cite{Schomerus1999,SW1999} to be
\beq
  \label{NCparameter}
  \theta^{IJ} = - g^{IK}\cF_{KL}G^{LJ}.
\eeq
On a coisotropic A-brane this bivector is non-trivial only transverse to the foliation. Note however that the corresponding noncommutative product will not be associative, since $\theta$ is generically not a Poisson structure \cite{CS2001,HKK2001}. 

However, in the case of a transverse hyper-K\"ahler structure, as discussed in Section~\ref{subsec:hyperKahler}, $\theta$ is indeed Poisson. To see this we apply (\ref{hyperKahler}) twice and obtain
$$
  \theta^{IJ} = \frac 12 \cF^{-1}{}^{IJ} \qquad \mathrm{on}~N\fol.
$$
In particular, it has type $(\widehat{2,0})\oplus(\widehat{0,2})$ with respect to $\cstr$ and using (\ref{HatDecomposition}) we find
\beq
  \label{hyperPoisson}
  \theta^{ab} = -\frac i2 \poisson^{ab},\qquad
  \theta^{\ba\bb} = \frac i2 \poisson^{\ba\bb}.
\eeq

In the following we do not want to restrict ourselves to the hyper-K\"ahler case though. Instead, to disentangle the noncommutative effects from other string dynamics we consider the topologically twisted theory, where the noncommutative geometry on the coisotropic A-brane will arise from deformation quantization \cite{CF1999}.%
\footnote{Notice that, according to Cattaneo and Felder \cite{CF2003}, coisotropic submanifolds (without transverse complex structure) also describe D-branes in the Poisson sigma-model. Although this theory is different from the topological A-model, one might learn some lessons from their work.
}
However, we do not want to go as far as summing up the Kontsevich product to all orders in $\theta$ \cite{Kontsevich1997}. In fact, we will only concentrate on the leading term in the noncommutative product.

In the topological A-model let us consider the trivial instanton sector with quantum fluctuations around the constant map into the target space $M$. Because of the boundary conditions the image of the constant map must lie on the coisotropic A-brane. In order to include the fluctuations in the path integral, we choose local coordinates $(x,\nu)$ near the coisotropic submanifold so that $\coiso$ lies at $\nu=0$. In these coordinates the action (\ref{topolaction}) becomes
\beq
  \label{expandtopolaction}
  S = \int_\Si (\om-i\cF)_{IJ} dx^I\!\wedge\!dx^J + 
     \int_\Si 2(\om-iB)_{IJ} d\nu^I\!\wedge\!dx^J +
     \{Q,V\} .
\eeq
Let us concentrate on the leading contribution to the noncommutative product. The second term in (\ref{expandtopolaction}) pairs $N\coiso$ with $T\fol$, and therefore can not act on topological observables, \ie it can only appear in higher order terms in the noncommuative product. On the other hand, the first term becomes
$$
  S = 2\int_\Si \om_{ab} dz^a\!\wedge\!dz^b + \cO(\nu) + \{Q,\ldots\}.
$$
On any two topological observables $f,g \in \hat H^0(\coiso)$ the noncommutative product therefore acts as
\beq
  \label{NCP}
  f * g = f g ~+~ \frac 14 \{f,g\}_\PB ~+~ \cO(\poisson^{2}),
\eeq
where $\{f,g\}_\PB$ is the Poisson bracket for $\poisson$,
$$
  \{f,g\}_\PB = \poisson^{ab} \dl_a f~ \dl_b g.
$$

The higher order terms in (\ref{NCP}) contain also contributions from the second part of the action (\ref{expandtopolaction}), which leads to terms in the Kontsevich product with derivatives normal to $\coiso$ \cite{CF2003}.
However, for space-filling coisotropic A-brane, $\coiso=M$, the noncommutative product (\ref{NCP}) will be the Kontsevich product for the holomorphic Poisson bivector $\pi^{ab}$.

The fact that we have $\theta^{ab}=\frac 12 \cF^{-1}{}^{ab} = -\frac i2\poisson^{ab}$ in the topological sector nicely ties in with the holomorphic Poisson structure (\ref{hyperPoisson}) in the transverse hyper-K\"ahler case. Note however that in the topological sector the leading term in the noncommutative product is governed by the Poisson structure $\theta = \frac 12 \cF^{-1}$ even if $C$ does not carry a transverse hyper-K\"ahler structure.

\section{Gauge fields and tachyons on coisotropic A-branes}
\label{AandT}

In this section we consider higher rank A-branes as well as stacks of branes and antibranes. We assume that the stack is supported on a coisotropic submanifold $C$ with a fixed transverse complex structure $\cstr$.
The gauge field $\hat A$, associated with a higher rank gauge group, and the tachyon profile $T$ will be treated as perturbations on the background that was studied in the previous section.
In the non-linear sigma model the data for a D-brane can be encoded in a superconnection $\cA$, whose path-ordered exponential is integrated along the world sheet boundary,
$$
  U(\s_i,\s_f) = P \exp\left(i\int_{\s_i}^{\s_f}\!\!\!dt \cA\right) .
$$
The integration is from an initial to a final point on the boundary, where boundary observables are  inserted.
The strategy of this section is to start with the general $N=1$ superconnection as in \cite{Hori2000,KL2000,TTU2000} and to determine the conditions on the gauge field and the tachyon profile from $N=2_A$ supersymmetry.

On general grounds, a line operator is invariant under an infinitesimal symmetry, say with parameter $\varepsilon$, if its variation leads to a total derivative, that is there exists a quantity $\cQ$ so that
$$
  \de U(\s_i,\s_f) =
  i\int_{\s_i}^{\s_f} \!\!\!dt ~U(\s_i,t)~ \de \cA(t)~ U(t,\s_f) \stackrel{!}{=}
  i\varepsilon\int_{\s_i}^{\s_f} dt ~d_t \left(U(\s_i,t) ~\cQ(t)~ U(t,\s_f)\right).
$$
Evaluating the right-hand side requires taking into account contact terms at the world sheet boundary,
$$
  i\varepsilon\int_{\s_i}^{\s_f} dt~  U(\s_i,t)\left( ~d_t\cQ(t) + 
  \lim_{\de t \rarrow 0} [\cA(t-\de t) \cdot \cQ(t) - \cQ(t) \cdot \cA(t+\de t)]
  \right)U(t,\s_f).
$$
Here the dot indicates the operator product of boundary operators. 
The so-called boundary charge $\cQ$ has the effect of twisting the infinitesimal symmetry on boundary operators,
$\de^{bdry} \cO := \de \cO+ i \varepsilon (\cQ \cdot \cO - \cO \cdot \cQ)$.
The invariance of the path-ordered exponential is then equivalent to the following condition on the superconnection,
\beq
  \label{bdrycharge}
  \de^{bdry} \cA = \varepsilon d_t \cQ \ .
\eeq

If the boundary conditions on a D-brane are of mixed von Neumann type, like in (\ref{N1conditions}), the contact terms will contain a contribution from the noncommutative product induced from the bivector $\theta$ \cite{Schomerus1999,SW1999}, These contact terms must be taken into account when we introduce the $N=1$ superconnection $\cA$ below. However, in general it is hard to disentangle the noncommutative effects from the stringy dynamics governed by the boundary metric $G$. In Sections~\ref{holomorphicgaugefield} and \ref{tachyon} we will therefore introduce the noncommutative product in the $N=1$ superconnection $\cA$ only formally. This can be done because in the method that we choose to check $N=2_A$ invariance the contact terms will not appear explicitly. Section~\ref{subsec:roleNCG} will then focus on the topologically twisted theory, where the contact terms are under control. As discussed in the previous section they are literally given by the holomorphic noncommutative product (\ref{NCP}).

\subsection{A holomorphic gauge field}
\label{holomorphicgaugefield}

Consider a stack of A-branes on a coisotropic submanifold $C$ with transverse complex structure $\cstr$. In this statement we already assumed to split off a part from the gauge field, which is proportional to the identity and enters into $\cstr$. The remaining non-Abelian gauge field, say $\hat A$, is treated perturbatively in an $N=1$ superconnection,
$$
  \cA = \hat A_I \dl_0 x^I - \frac i4 \hat F_{IJ}\psi^I  \psi^J .
$$
Here, the curvature of the gauge field $\hat A$ is
$$
  \hat F_{IJ} = \dl_I \hat A_J - \dl_J \hat A_I + 
  i (\hat A_I * \hat A_J - \hat A_J * \hat A_I),
$$
where the star denotes the noncommutative product.

Instead of checking the invariance under $N=2_A$ supersymmetry (\ref{NAalgebra}) directly, which would involve taking care of complicated contact terms, we take an easier route: Since the $N=2_A$ supercurrents are an axial R-symmetry rotation of the $N=1$ currents, we just need to check invariance under axial R-symmetry. The $N=2_A$ invariance is then automatic. To this end we rewrite the second term in the superconnection in terms of the A-model coordinates, that is 
$\hat F_{IJ}\psi^I  \psi^J = \hat F_{IJ} (\chi^I\chi^J + 2 \chi^I\rho^J + \rho^I \rho^J)$. Because of (\ref{N2Acoordinates}), the first and last term are not invariant and thus put constraints on the allowed curvature $\hat F$. Recall that the boundary conditions (\ref{N2Afermcondition}) and (\ref{normalfermcondition}) state that $\chi$  vanishes on covectors in $N^{(\widehat{1,0})}\fol^*$ and that $\rho$ vanishes on covectors in $N^{(\widehat{0,1})}\fol^*$. This implies the following: First, the curvature $\hat F$ has to be transverse to the foliation, \ie for all $t \in T\fol$ we have $t^I \hat F_{IJ} = 0$. 
Second, the curvature $F$ must be of type $(\widehat{1,1})$ with respect to the transverse complex structure $\cstr$. The superconnection then becomes
\beq
  \label{superconnect}
  \cA = \hat A_I \dl_0 x^I - \frac i2 \hat F_{a\bb}~\rho^a \chi^\bb . 
\eeq
We found that:

\
 
\parbox{15cm}{
\emph{A higher rank A-brane on a coisotropic submanifold $C$ with transverse complex structure $\cstr$ is a noncommutative transversely holomorphic (nth) vector bundle $E$ \cite{Gomez1980}, which we define as a vector bundle equipped with a connection $\hat A$, whose curvature vanishes along the leaves of the characteristic foliation and is noncommutative and of type $(\widehat{1,1})$ in transverse direction.}
}

\

\noindent Algebraically, such an A-brane can be characterized by the sheaf $\cE_C$ of analytic sections of the nth-vector bundle $E$, \ie the sheaf of analytic sections that are noncommutative transversely holomorphic \cite{Gomez1980}. We will refer to such an object as nth-sheaf.

In view of these results the rank one coisotropic A-brane from the previous section is a coisotropic submanifold $C$ with transverse complex structure $\cstr$ and a trivial nth-line bundle $L$. Its sheaf of analytic sections is the transverse structure sheaf $\fol\cO_{C}$ on $C$, that is the sheaf of functions that are noncommutative transversely holomorphic. The sheaf $\cE_C$ representing a higher rank A-brane on $C$ is then a locally free left-$\fol\cO_{C}$-module \cite{Gomez1980}.

\subsubsection*{Some remarks on the Seiberg--Witten map}

So far we considered stacks of A-branes with fixed transverse complex structure $\cstr$. In order to compare two A-branes, still on the same coisotropic submanifold, but with different complex structures, we will now apply the Seiberg--Witten map \cite{SW1999}.
To keep control we will require that the bivector for the noncommutative product is given by $\theta = 1/2 \cF^{-1} = -1/2 \cstr \pi$. The following arguments will therefore be true for the topologically twisted theory.

In general, the Seiberg--Witten map is too complicated to be useful for our purposes. However, in flat space, say on the torus, and for constant \emph{Abelian} (but possibly higher rank) gauge fields
it simplifies considerably: The map between the curvature $\hat F$ of the noncommutative connection and the curvature $\tilde F$ of the commutative one is given by \cite{SW1999},
\beq
  \label{SWmap}
  \hat F ~=~ \left(\id+ \tilde F \theta\right)^{-1} \tilde F ~=~ 
             \tilde F \left(\id+ \theta\tilde F \right)^{-1} 
             \qquad \mathrm{on}\quad N\fol.
\eeq
Applying this transformation to the condition that $\hat F$ is of type $(\widehat{1,1})$,
that is
\beq
  \label{type11}
  \cstr^t \hat F + \hat F \cstr = 0 \qquad \mathrm{on}\quad N\fol,
\eeq
gives the equation
$$
  \cstr^t \tilde F + \tilde F \cstr + \tilde F \pi \tilde F = 0 \qquad \mathrm{on}\quad N\fol.
$$
The commutative curvature can now be combined with the background into 
$F_{tot} = \cF + \tilde F$, on which condition (\ref{type11}) becomes
\beq
  \label{Ftotal}
  \pi F_{tot}~ \pi F_{tot} = - \id_{N\fol} \otimes \id_E .
\eeq
For rank one gauge fields, $F_{tot}$ defines a new transverse complex structure on $C$. For a higher rank Abelian gauge field, $F_{tot}$ defines a stack of rank one coisotropic A-branes on $C$, possibly with different transverse complex structures.

Therefore, in the constant and Abelian case the Seiberg--Witten map can be used to map a coisotropic A-brane with $\cstr_1$ and curvature $\hat F_1$ to one with another transverse complex structure $\cstr_2$. In fact, two Abelian stacks of coisotropic A-branes, say $(C,\cstr_1,\hat F_1)$ and $(C,\cstr_2,\hat F_2)$, are equivalent if and only if their curvatures satisfy the field redefinition
\beq
  \label{composeSWmap}
\left(\cF_1 +\frac 12 \hat F_1\right)\left(\cF_1 -\frac 12 \hat F_1\right)^{-1}\!\cF_1 ~=~ 
\left(\cF_2 +\frac 12 \hat F_2\right)\left(\cF_2 -\frac 12 \hat F_2\right)^{-1}\!\cF_2.
\eeq

From the above discussion it is clear that this relation preserves the condition that the noncommutative curvature is of type $(\widehat{1,1})$. 
It is natural to expect that the analogous map for the general non-Abelian case also preserves this property. An argument in favour of this expectation goes as follows. The Seiberg--Witten map is merely a field redefinition that relates the fields coming from different choices of the regularization scheme of the two-dimensional quantum field theory. Such field redefinitions do not change the S-matrix and, in particular, preserve supersymmetry. 
So, if we apply a field redefinition --- the analog of (\ref{composeSWmap}) --- on an A-brane $(C,\cstr_1,\hat F_1)$, the resulting A-brane $(C,\cstr_2,\hat F_2)$ must be supersymmetric, and therefore $\hat F_2$ must have type $(\widehat{1,1})$.
For a non-Abelian gauge group it is however unclear how the commutative formulation --- the analog of relation (\ref{Ftotal}) --- should look like.

If this expectation is indeed true and if a coisotropic submanifold $C$ admits several transverse complex structures, we can always pick one of them, say $\cstr$. Every higher rank coisotropic A-brane on $C$ can then be written in terms of the superconnection (\ref{superconnect}) for some nth-vector bundle $E$.

\subsection{A holomorphic tachyon profile}
\label{tachyon}

Next, we consider brane--antibrane pairs with tachyon profile. 
Given a vector bundle $E$ over $C$ a tachyon profile is an endomorphism 
$T \in \End(E)$, which is hermitian with respect to a hermitian metric on $E$. The corresponding $N=1$ superconnection is given by \cite{Hori2000,KL2000,TTU2000,HHP2008}
\beq
  \label{N1tachyon}
  \cA = \frac i2 \psi^I \hat D_I T + \frac 12 T*T .
\eeq
Here, $\hat D T = d T + i (A*T - T*A)$ denotes the covariant derivative. For subsequent reference we split it according to the foliation $\fol$, $\hat D = d_{\parallel,\hat A} + \dl_{\perp,\hat A} + \bar\dl_{\perp,\hat A}$, and introduce 
$$
  d_{C,\hat A} = d_{\parallel,\hat A} +\bar\dl_{\perp,\hat A}.
$$

To check axial R-symmetry we rewrite the first term in (\ref{N1tachyon}) as 
$i/2(\rho^I \hat D_I T + \chi^I \hat D_I T)$. The only way to make it R-invariant is to split the tachyon profile into $T = T_+ + T_-$, where the two parts transform homogeneously with charges $\pm 1$ under R-rotations, that is for a given representation $R(\la)$ of the R-symmetry group on the vector bundle $E$ we have
$$
  R(\la) T_\pm(x) R(\la)^{-1} = \la^{\pm 1} T_\pm(x).
$$
Taking into account the boundary conditions (\ref{N2Afermcondition}) and (\ref{normalfermcondition}) for the fermions $\chi$ and $\rho$ we therefore need that $d_{\parallel,\hat A}T_\pm=0$ and $\dl_{\perp,\hat A} T_+ = 0 = \bar\dl_{\perp,\hat A} T_-$.
Let us introduce the notation $T_- = i Q$ for the transversely holomorphic part of the tachyon profile. By the hermiticity of $T$ we have $T_+ = - i Q^\dagger$. The coordinate dependence of $Q$ can be summarized in
\beq
  \label{holomorphictachyon}
  d_{C,\hat A} Q = 0\ .
\eeq
$Q$ is therefore an endomorphism on an nth-vector bundle $E$. 

In order to see the consequence of axial R-invariance of the second term in (\ref{N1tachyon}), we write it as
$T*T = -Q*Q + \{Q\stackrel{*}{,}Q^\dagger\} - Q^\dagger * Q^\dagger$. It is invariant if and only if
\beq
  \label{Qdiff}
  Q*Q = 0,
\eeq
that is if $Q$ is a noncommutative transversely holomorphic differential on $E$.
Combining results we obtain the following $N=2_A$ superconnection,
$$
  \cA = \hat A_I \dl_0 x^I - \frac i2 \hat F_{a\bb}\rho^a  \chi^\bb
        -\frac 12 \chi^a \hat D_a Q + \frac 12 \rho^\ba \hat D_\ba Q^\dag
      + \frac 12 \{Q\stackrel{*}{,}Q^\dag\} .
$$

The action of the R-symmetry group on $Q$ is given by
\beq
  \label{Rcharge1}
  R(\la) Q(z) R(\la)^{-1} = \la Q(z),
\eeq
that is $Q$ has R-degree $1$. If we assume $Q$ to be irreducible, it therefore splits the vector bundle into a 
finite direct sum, $E = \oplus_{j} E^j$, where $j \in \bbZ$ denotes the R-degree.
Representing the vector bundles $E^j$ over $C$ by 
nth-sheaves $\cE^j_C$, the conditions (\ref{holomorphictachyon}--\ref{Rcharge1}) on the tachyon profile $Q$ say that the A-brane is determined by a complex of locally free left-$\fol\cO_C$-modules,
\beq
  \label{complex}
  \ldots ~\mapshort{Q}~ \cE^{j-1}_C ~\mapshort{Q}~ \cE^{j}_C ~\mapshort{Q}~ \cE^{j+1}_C ~\mapshort{Q}~ \ldots\ .
\eeq

The results of this section can be summarized as follows:

\

\parbox{15cm}{
\emph{An A-brane on a coisotropic submanifold $C$ with fixed transverse complex structure
  $\cstr$ is a complex (\ref{complex}) of nth-sheaves (vector bundles) with noncommutative
  transversely holomorphic differential $Q$.
}}

\subsection{The role of noncommutative geometry for tachyon condensation}
\label{subsec:roleNCG}

We now perform the topological A-twist in order to get control over the noncommutative product. In the twisted theory the conditions (\ref{holomorphictachyon}--\ref{Rcharge1}) for the A-brane are unchanged, but the noncommutative product therein is now given explicitly by (\ref{NCP}), which is governed by the holomorphic bivector 
$\theta^{ab} = -i/2 \pi^{ab}$.

In order to understand the role of noncommutative geometry for the tachyon condensation process let us consider a special complex, namely the Koszul complex for an $N$-tuple of sections of nth-line bundles, $\underline{f}=\{f_1,\ldots,f_N\}$. We denote the associated nth-sheaves by $\cL_{C,l}$ for $l = 1,\ldots, N$. By setting 
$\cV_C = \oplus_l \cL^{-1}_{C,l}$ the Koszul complex is given by
$$
  \wedge^{N}\!\cV_C ~~\mapshort{\underline{f}}~
  \wedge^{N-1}\!\cV_C ~\mapshort{\underline{f}}~ 
  \ldots ~\mapshort{\underline{f}}~
  \wedge^1\!\cV_C 
  ~\mapshort{\underline{f}}~ \fol\cO_C .
$$

It will localize in the infra-red on the zero locus $\cap_l\{f_l=0\}$, which in case of a complete intersection will have complex codimension $N$. At first sight, to have a non-trivial zero locus requires the number of sections to be less or equal to 
$\dim N\fol^{(\widehat{1,0})} = 2k$. However, this bound is too big, as in the maximal case this would lead to a submanifold of real dimension $n-2k$, which is less than the dimension of a Lagrangian submanifold and thus inconsistent with the observation that A-branes must be supported on coisotropic submanifolds.

The resolution to this problem is provided by the property that $Q$ is a noncommutative differential, $Q*Q=0$. Indeed, on the sections $\underline{f}$ this condition becomes
$$
  f_l * f_m - f_m * f_l = 0 \tfor m,l = 1,\ldots,N\ .
$$
Let us pick the unique connection for the bundle $E$ with 
$\bar \dl_{\perp,\hat A} = \bar\partial_\perp$. In this frame the sections $f_l$ depend holomorphically on the transverse coordinates $z^a$, and in leading order of the expansion (\ref{NCP}), the condition on the sections becomes
\beq
  \label{noncommQ2}
  \{f_l, f_m\}_\PB + \cO(\poisson^2) = 0 \tfor m,l = 1,\ldots,N\ ,
\eeq
\ie the sections $\underline{f}$ must be in involution. 
However, the maximal number of Poisson commuting sections for a rank $k$ holomorphic Poisson structure $\poisson$ is $k$, so that the maximal number of sections in the Koszul complex is $k$ (and not $2k$).
Indeed, condition (\ref{noncommQ2}) ensures that the zero locus $\cap_l\{f_l=0\}$ is again coisotropic.

Notice that with the help of Koszul complexes we can construct Lagrangian submanifolds by finding a maximal number of commuting sections $\underline{f}$. However, the converse is not true. Because of the required transverse complex structure on the coisotropic A-brane we can not in general find a Koszul complex for a given Lagrangian submanifold.

It is unclear at present, how the Seiberg--Witten map acts on complexes of nth-bundles. While we learned, at least in flat space with constant $\theta$, how the Seiberg--Witten map (\ref{composeSWmap}) acts on the connections of the line bundles in the Koszul complex, we need to understand how it transforms the differential of the complex. We would however expect two seemingly contradicting properties from such a transformation. First, since the transverse complex structure is changed under the Seiberg--Witten map, the transformed differential must be holomorphic with respect to the new transverse complex structure. Second, the zero loci associated with these two Koszul complexes, for instance Lagrangian submanifolds, should coincide. However, the zero loci of two sets of sections that are holomorphic with respect to two different complex structures cannot agree in general. At present, we can only speculate that a possible way out is to relax the second point and admit Hamiltonian deformations between the two zero loci.

\subsection{The topological observables}

In the topologically twisted theory the BRST operator is represented on the zero mode sector by a transversely holomorphic version of Quillens superconnection \cite{KL2000,TTU2000,Quillen1985}. It acts on the bundle $E = \oplus_j E^j$ by
$$
  \bar\nabla := d_{C,\hat A} - Q.
$$
This can be checked using the A-type boundary conditions and computing the boundary supercharge from (\ref{bdrycharge}) for the BRST transformations (\ref{topalgebra}). The conditions (\ref{holomorphictachyon}--\ref{Rcharge1}) turn Quillens superconnection into a  differential,
$$
  \bar\nabla * \bar\nabla = 0.
$$
The topological observables between two complexes are then represented by $\bar\nabla$-cohomology classes of forms in 
$$
  \hat\Omega^q(\coiso,Hom^*(E_1,E_2)) :=   
  \bigoplus_{n+m=q}\hat\Omega^n(\coiso,Hom^m(E_1,E_2)),
$$
where the form degree and the homological degree of the complex combine into the axial R-charge $q=n+m$. More precisely, the differential acts as
$$
  \begin{array}{ccccc}
  \nn
  \bar\nabla &:& \hat\Omega^q(\coiso,Hom^*(E_1,E_2)) 
  &\rightarrow&
  \hat\Omega^{q+1}(\coiso,Hom^*(E_1,E_2))\ ,\\[5pt]
  \nn
  && \omega_{21} &\mapsto& \bar\nabla_2*\omega_{21} - 
  (-1)^q~ \omega_{21} * \bar\nabla_1\ .
\end{array}
$$
Particular examples of elements in the cohomology group 
$\hat\cH^0(\coiso,Hom^*(E_1,E_2))$ are chain maps between complexes up to homotopy, \ie transversely holomorphic elements $\psi_{21} \in Hom^0(E_1,E_2)$ that are $Q$-closed, 
$Q_2 * \psi_{21} - \psi_{21} * Q_1 = 0$, up to $Q$-exact terms.

\section{Concluding remarks and challenges}
\label{sec:lessons}

\newcommand{\rb}[1]{\raisebox{6pt}[-6pt]{#1}}

Let us close with some lessons from this work and open problems for future research.
The geometric structures on coisotropic A-branes can be summarized in a table stressing the interplay between perturbative and non-perturbative geometric structures:

\begin{center}
\begin{tabular}{|c|c|c|}
  \hline
  && \\
  & \rb{On leaves of the foliation} & \rb{Transverse to the foliation}  \\
  \hline\hline
  \begin{sideways}
    \!\!\!\!\!\!\!\!\!\!\!\!\!\!\!\!\!\!\!\!\!\!\!\!\!\!\!\!\!\!\!\!\!\!\!\!
	  perturbative~~\ 
	\end{sideways}
  &  & complex structure:\\
  & \rb{------}& $\cstr^2 = -\id$  \\
  \cline{2-3} 
  &  & noncommutative product:\\[-2pt]
  & \rb{------} & 	from $(\widehat{2,0})$-bivector $\pi^{ab}$\\
  \cline{2-3} && tachyon profile: \\
  &\rb{------}& holomorphic, $Q * Q = 0$  \\
  \hline
  & gauge field: & noncommutative gauge field:   \\
  &$\hat F = 0$ & $\hat F$ of type $(\widehat{1,1})$ \\
  \hline
  \begin{sideways}\!\!\!\!\!\!\!\!\!\!\!\!\!\!\!\!\!\!\!\!non-pert.~~\ \end{sideways}&  disk instantons: & \\
  & with boundary along the & \\
  & leaves and weighted by the & \rb{------}\\
  & holonomy of the gauge field&\\
  \hline
\end{tabular}
\end{center}

In particular, there are two limiting cases of coisotropic A-branes, the ones with minimal dimension and the ones with maximal dimension. A-branes of the first sort are the familiar Lagrangian submanifolds, that is with no transverse direction to the foliation and hence no perturbative geometric structure. The non-trivial geometry comes solely from non-perturbative disk instanton corrections which measure the holonomy of the flat gauge connection.

The second limiting situation are space-filling coisotropic A-branes, $C=M$, which exist only when $M$ is complex even-dimensional. The geometric structure on space-filling A-branes is purely perturbative and resembles that of a space-filling B-brane, it is a $\cstr$-holomorphic vector bundles, albeit noncommutative. There are no instanton corrections. (Notice however that for complex odd-dimensional $M$ the coisotropic A-branes with maximal dimension may have instanton corrections due to their one-dimensional leaves.)

\subsubsection*{Noncommutative geometry vs. instanton corrections}

As we saw in the previous section tachyon condensation provides a means to construct lower-dimensional coisotropic A-branes, in particular Lagrangian ones, from maximal dimensional ones. It may therefore provide a direct link between disk instanton corrections of the former and noncommutative geometry of the latter. 

To make this more concrete, consider the four-torus $T^4$. Pick three Lagrangian submanifolds, $L_i$ for $i=1,2,3$, with flat connections and assume that they are the zero loci of global sections $\Theta_i$ of nth-line bundles $\cL_i$ over $T^4$. According to our discussion on tachyon condensation in the previous section we can represent these Lagrangian A-branes by two-term complexes of coisotropic A-branes (possibly tensored with a line bundle),
$$
  \cL_i^{-1} ~\mapshort{\Theta_i}~ \fol\cO.
$$

There are then two ways of computing the corresponding topological 3-point function on the disk.
\begin{itemize}
  \item The first involves explicit instanton counting of $J$-holomorphic discs, that is 
        triangles weighted 
        by the exponential of the complexified K\"ahler class, $q \sim e^{2 \pi i t}$. 
  \item The second uses algebraic methods over the noncommutative four-torus, with
        $u$ and $v$ being the $\cstr$-compex coordinates with commutation relations
        $[u,v] = \pi^{uv}$. For the exponentials, $U = e^{2\pi i u}$ and $V=e^{2\pi i v}$,
        we have $U V = q' V U$ with $q' = e^{2 \pi i\,\pi^{uv}} \sim e^{2\pi i / t}$. 
\end{itemize}
The expectation is that the first computation leads to a theta function depending on the modular parameter $q$ and the second gives a theta function depending on $q'$. These should then be related by a modular transformation.

It would interesting to work out this connection between noncommutative geometry and instanton counting in more detail. However, fixing all the details in the outline above is out of the scope of this paper and will be treated elsewhere.

\subsubsection*{Open problems on the route to a category of A-branes}

In this work we only considered complexes of nth-vector bundles on a single coisotropic submanifold $C$ with fixed transverse complex structure $\cstr$. However, the most general A-brane will be a complex that involves objects that are holomorphic with respect to different transverse complex structures and that are supported on different coisotropic submanifolds.

Such complexes require an understanding of
the topological observables (of R-degree $0$) between two nth-vector bundles, say $E_1$ over $(C_1, \cstr_1)$ and $E_2$ over $(C_2,\cstr_2)$, as they will be the morphisms that build up the differentials in general complexes. However, already for $C_1=C_2$ the BRST operator is represented by two different Dolbeault operators, one associated with $\cstr_1$ the other with $\cstr_2$. What is the meaning of the holomorphicity condition, $d_{C,\hat A} Q = 0$, in such a situation?

Having settled this issue, the next one is to form a differential that satisfies the analog of the relation $Q*Q=0$. However, in order to write down such a relation we first need to understand how to compose morphisms between nth-vector bundles. Here, a familiar issue with A-branes drops in, the composition will in general be a quantum product that includes instanton corrections.

\begingroup\raggedright

\endgroup


\begin{thebibliography}{10}

\bibitem{KO2001}
  A.~Kapustin and D.~Orlov,
  ``Remarks on A-branes, mirror symmetry, and the Fukaya category,''
  J.\ Geom.\ Phys.\  {\bf 48} (2003) 84
  [arXiv:hep-th/0109098].

\bibitem{KW2006}
  A.~Kapustin and E.~Witten,
  ``Electric-magnetic duality and the geometric Langlands program,''
  arXiv:hep-th/0604151.

\bibitem{GW2008}
  S.~Gukov and E.~Witten,
  ``Branes and Quantization,''
  arXiv:0809.0305 [hep-th].

\bibitem{NW2010}
  N.~Nekrasov and E.~Witten,
  ``The Omega Deformation, Branes, Integrability, and Liouville Theory,''
  arXiv:1002.0888 [hep-th].

%\bibitem{Zabzine2004}
%  M.~Zabzine,
%  ``Geometry of D-branes for general N = (2,2) sigma models,''
%  Lett.\ Math.\ Phys.\  {\bf 70} (2004) 211
%  [arXiv:hep-th/0405240].
%
%\bibitem{GM2004}
%  P.~Grange and R.~Minasian,
%  ``Modified pure spinors and mirror symmetry,''
%  Nucl.\ Phys.\  B {\bf 732} (2006) 366
%  [arXiv:hep-th/0412086].

\bibitem{KL2003}
  A.~Kapustin and Y.~Li,
  ``Stability conditions for topological D-branes: A worldsheet approach,''
  arXiv:hep-th/0311101.

\bibitem{Li2004}
  Y.~Li,
  ``Anomalies and graded coisotropic branes,''
  JHEP {\bf 0603} (2006) 100
  [arXiv:hep-th/0405280].

\bibitem{KO2003}
  A.~Kapustin and D.~Orlov,
  ``Lectures on mirror symmetry, derived categories, and D-branes,''
  arXiv:math/0308173.

\bibitem{KL2002}
  A.~Kapustin and Y.~Li,
  ``D-Branes in Landau-Ginzburg Models and Algebraic Geometry,''
  JHEP {\bf 0312} (2003) 005
  [arXiv:hep-th/0210296].

\bibitem{BHLS2003}
  I.~Brunner, M.~Herbst, W.~Lerche and B.~Scheuner,
  ``Landau-Ginzburg realization of open string TFT,''
  JHEP {\bf 0611}, 043 (2006)
  [arXiv:hep-th/0305133].

\bibitem{Lazaroiu2003}
  C.~I.~Lazaroiu,
  ``On the boundary coupling of topological Landau-Ginzburg models,''
  JHEP {\bf 0505}, 037 (2005)
  [arXiv:hep-th/0312286].

\bibitem{Hori2000}
  K.~Hori,
  ``Linear models of supersymmetric D-branes,''
  arXiv:hep-th/0012179.

\bibitem{HHP2008}
  M.~Herbst, K.~Hori and D.~Page,
  ``Phases Of N=2 Theories In 1+1 Dimensions With Boundary,''
  arXiv:0803.2045 [hep-th].

\bibitem{FIM2006}
  A.~Font, L.~E.~Ibanez and F.~Marchesano,
  ``Coisotropic D8-branes and model-building,''
  JHEP {\bf 0609} (2006) 080
  [arXiv:hep-th/0607219].

\bibitem{HW2004}
  K.~Hori and J.~Walcher,
  ``F-term equations near Gepner points,''
  JHEP {\bf 0501} (2005) 008
  [arXiv:hep-th/0404196].

\bibitem{Kapustin2003}
  A.~Kapustin,
  ``Topological strings on noncommutative manifolds,''
  Int.\ J.\ Geom.\ Meth.\ Mod.\ Phys.\  {\bf 1} (2004) 49
  [arXiv:hep-th/0310057].

\bibitem{Kapustin2005}
  A.~Kapustin,
  ``A-branes and noncommutative geometry,''
  arXiv:hep-th/0502212.

\bibitem{KL2000}
  P.~Kraus and F.~Larsen,
  ``Boundary string field theory of the DD-bar system,''
  Phys.\ Rev.\  D {\bf 63} (2001) 106004
  [arXiv:hep-th/0012198].

\bibitem{TTU2000}
  T.~Takayanagi, S.~Terashima and T.~Uesugi,
  ``Brane-antibrane action from boundary string field theory,''
  JHEP {\bf 0103} (2001) 019
  [arXiv:hep-th/0012210].

\bibitem{BS2002}
  P.~Bressler and Y.~Soibelman,
  ``Mirror symmetry and deformation quantization,''
  arXiv:hep-th/0202128.

\bibitem{SSW2009}
  A.~Sevrin, W.~Staessens and A.~Wijns,
  ``An N=2 worldsheet approach to D-branes in bihermitian geometries: II. The
  general case,''
  JHEP {\bf 0909} (2009) 105
  [arXiv:0908.2756 [hep-th]].

\bibitem{MirrorBook}
  K.~Hori {\it et al.},
  ``Mirror symmetry,''
%\href{/spires/find/hep/www?irn=6746918}{SPIRES entry}
{\it  Providence, USA: AMS (2003) 929 p}

\bibitem{HIV2000}
  K.~Hori, A.~Iqbal and C.~Vafa,
  ``D-branes and mirror symmetry,''
  arXiv:hep-th/0005247.

\bibitem{Oh2003}
  Y.~G.~Oh,
  ``Geometry of coisotropic submanifolds in symplectic and Kähler manifolds,'' 
  arXiv:math.SG/0310482.

\bibitem{SW1999}
  N.~Seiberg and E.~Witten,
  ``String theory and noncommutative geometry,''
  JHEP {\bf 9909}, 032 (1999)
  [arXiv:hep-th/9908142].

\bibitem{KL2005}
  A.~Kapustin and Y.~Li,
  ``Open string BRST cohomology for generalized complex branes,''
  Adv.\ Theor.\ Math.\ Phys.\  {\bf 9} (2005) 559
  [arXiv:hep-th/0501071].

\bibitem{KM2006}
  P.~Koerber and L.~Martucci,
  ``Deformations of calibrated D-branes in flux generalized complex
  manifolds,''
  JHEP {\bf 0612} (2006) 062
  [arXiv:hep-th/0610044].

\bibitem{OP2003}
  Y.~G.~Oh and J.~S.~Park,
  ``Deformations of coisotropic submanifolds and strongly homotopy Lie
  algebroids,''
  Invent. Math. {\bf 161} (2005) 287
  arXiv:math/0305292.

\bibitem{DK1979}
  T.~Duchamp, M.~Kalka,
  ``Deformation theory for holomorphic foliations,''
  J. Diff. Geometry {\bf 14}, 317 (1979)

\bibitem{Gomez1980}
  X.~Gomez--Mont,
  ``Transversal holomorphic structures,''
  J. Diff. Geometry {\bf 15}, 161 (1980)

\bibitem{Schomerus1999}
  V.~Schomerus,
  ``D-branes and deformation quantization,''
  JHEP {\bf 9906} (1999) 030
  [arXiv:hep-th/9903205].

\bibitem{CS2001}
  L.~Cornalba and R.~Schiappa,
  ``Nonassociative star product deformations for D-brane worldvolumes in
  curved backgrounds,''
  Commun.\ Math.\ Phys.\  {\bf 225} (2002) 33
  [arXiv:hep-th/0101219].

\bibitem{HKK2001}
  M.~Herbst, A.~Kling and M.~Kreuzer,
  ``Star products from open strings in curved backgrounds,''
  JHEP {\bf 0109} (2001) 014
  [arXiv:hep-th/0106159].

\bibitem{CF1999}
  A.~S.~Cattaneo and G.~Felder,
  ``A path integral approach to the Kontsevich quantization formula,''
  Commun.\ Math.\ Phys.\  {\bf 212} (2000) 591
  [arXiv:math/9902090].

\bibitem{CF2003}
  A.~S.~Cattaneo and G.~Felder,
  ``Coisotropic submanifolds in Poisson geometry and branes in the Poisson
  sigma model,''
  Lett.\ Math.\ Phys.\  {\bf 69} (2004) 157
  [arXiv:math/0309180].

\bibitem{Kontsevich1997}
  M.~Kontsevich,
  ``Deformation quantization of Poisson manifolds, I,''
  Lett.\ Math.\ Phys.\  {\bf 66}, 157 (2003)
  [arXiv:q-alg/9709040].
  
\bibitem{Quillen1985}
D.~Quillen,
``Superconnections and the Chern character,''
 Topology~{\bf 24} (1985) 89-95.


  


\end{thebibliography}
\end{document}